\documentclass{desyproc}
\def\bb{\begin{equation}}
\def\ee{\end{equation}}
\def\ba{\begin{eqnarray}}
\def\ea{\end{eqnarray}}
\def \eqf{\begin{eqnarray}}
\def \eea{\end{eqnarray}}
\def \eqb{\begin{eqnarray}}
\def \eqf{\end{eqnarray}}
\def \ee{\end{equation}}
\def \be{\begin{equation}}
\def \bea{\begin{eqnarray}}
\def \eea{\end{eqnarray}}

\begin{document}
\title{Aharonov-Bohm Effect and Hidden Photons}

\author{{\slshape Paola Arias}\\[1ex]
Pontificia Universidad Catolica de Chile, Casilla 306, Santiago 22, Chile.}
\contribID{familyname\_firstname}

\desyproc{DESY-PROC-2013-XX}
\acronym{Patras 2013} 
\doi  

\maketitle

\begin{abstract}
Signs of hypothetical light gauge bosons from a hidden sector may appear in Aharonov-Bohm-like experiments. The absence of signal in carried on experiments allow us to set a modest constraint to the mass and coupling constant of these particles. Our findings open the possibility to exploit the leaking of {\it hidden magnetic field} in a different setup of experiments.
\end{abstract}
\section{Introduction}
Hidden sectors are  on vogue nowadays since they are needed to realize  popular extensions of the Standard Model, as Supersymmetry and String Theory. Hidden sectors could be very weakly coupled to our visible sector through heavy messengers, via loop interactions. As a consequence, we are left with effective couplings at low energies, that mix both sectors. Prime candidates, that can emerge from a hidden sector, are hidden photons (HP), i.e., gauge bosons of an extra U(1) gauge group \cite{Abel:2008ai}. These particles, if massive, mix kinetically with photons, leading to oscillations between both of them.

Several direct and indirect searches have been performed looking for these particles and many others are being planned for the near future (see ref.\cite{Ringwald:2012hr} and references therein). Even though HPs are very weakly coupled to photons, they do leave an imprint in several physical phenomena, for instance, Coulomb law \cite{Bartlett:1988yy}. In the same way, we want to explore signatures the Aharonov-Bohm (AB) effect \cite{Aharonov:1959fk}, can gives us about hidden photons.
\section{Aharonov Bohm effect for massive photons}
The observable essence of the AB effect is the path-dependent phase, $\varphi$, of an electron wavefunction, 
which is shifted in the presence of an electromagnetic potential
\be
\exp \left(i e \oint_c \vec A \cdot d\vec x \right)\equiv \exp (i \Delta \varphi) \quad.
\ee
Where the phase shift $\Delta \varphi$ is related with the magnetic flux, $\Phi$, enclosed by the path of the electron, $\Delta \varphi=e\Phi$. 
A theoretical modification of the AB effect under a possible  non-zero photon mass was discussed in Boulder and Deser (BD), \cite{Boulware:1989up}. They argue that it is not necessary for the vector potential to be a gauge field achieve the AB effect, but to be minimally coupled to matter. 

Let us assume a typical AB experiment, with a solenoid of radius $a$, and a steady current,  $j$. Following BD, for a massive photon, the equation of motion of the system is the Proca equation. For stationary currents, only the transverse mode survives and the equation of motions becomes
\bb 
\left(-\nabla^2+M^2\right) {\bf A}=e{\bf  J}. \label{mastereq}
\ee
Where $M$ is the photon mass.The current in a magnet solenoid has the form ${\bf{ J}}=\hat \varphi j\delta(a-\rho)$. Where $\rho$ and $\varphi$ are cylindrical coordinates.

Imposing a cylindrically symmetric ansatz for the vector potential, of the form ${\bf{A}}=\hat z\times \nabla \Pi(\rho)$ we find an equation for $\Pi(\rho)$
\bb
\left(\partial_\rho^2+ \rho^{-1} \partial_\rho -M^2\right)\Pi(\rho)=ej\Theta(a-\rho).
\ee
Whose solution is given by 
\bb
\footnotesize{
\Pi(\rho)=-j e \left[ -\theta(\rho-a) K_0(m\rho) \int_0^a \rho' d\rho' I_0(m\rho')+\theta(a-\rho)\left( K_0(m\rho)\int^\rho_0 \rho' d\rho' I_0(m\rho')+I_0(m\rho) \int_\rho^a \rho'd\rho' K_0(m\rho')\right)\right].}
\label{pieq}
\ee
The corresponding magnetic field is given by the expression ${\bf{B}}=\nabla\times {\bf{A}}$. By taking the curl of the vector potential, we get ${\bf{B}}=\hat z\, e\, j \Theta(a-\rho)+\hat zM^2\Pi(\rho).$

We note that besides the usual solution -- confined to the surface of the solenoid -- there is a contribution that leaks out with a range given by $M^{-1}$. Therefore, the magnetic flux besides its normal value, picks up extra contributions from inside and outside the solenoid, coming from the non-zero mass of the photon.


\section{Aharonov Bohm effect for hidden photons}
The effective low energy Lagrangian that mixes photons, $A_\mu$, with hidden photons,  $X_\mu$, is
\be\label{Lagrange0}
{\mathcal{L}}=-\frac{1}{4} F_{\mu \nu} F^{\mu \nu}- \frac{1}{4} G_{\mu \nu} G^{\mu \nu}+
\frac{\sin \chi}{2} G_{\mu \nu} F^{\mu \nu}+ \frac{\cos^2 \chi}{2} m_{\gamma'}^2 X_\mu X^\nu+ J_\mu A^\mu\quad,
\ee
where $F_{\mu\nu}$ is the field strength tensor of photons and $G_{\mu\nu}$ the analogue for hidden photons.
The quantity $\chi$ accounts for the strength of the coupling between visible and hidden sectors and is predicted to be very small \cite{Dienes:1996zr}.
We have also included a mass term for the hidden photon, $m_{\gamma'}$, arising from a standard Higgs mechanism or Stueckelberg mechanism \cite{Ahlers:2008qc}.

Kinetic mixing can be removed from the Lagrangian by rotating the fields to a new basis, with a massless photon and a heavy hidden photon, {\it{i.e.}} $\tilde B_\mu  = B_\mu \cos \chi$, and  $\tilde A_\mu = A_\mu - \sin \chi B_\mu $

In this new basis the equations of motion for (\ref{Lagrange0}) read
\eqb
-\nabla^2 {\bf{\tilde A}}&=& e~{\bf{J}},\\
\left(-\nabla^2+m_{\gamma'}^2\right){\bf{\tilde X}}&=&e \tan\chi {\bf{J}}.
\eqf
Where we have chosen a gauge such $\tilde A_0=\tilde X_0=0$. The equation of motion for the field ${\bf{\tilde A}}$ is the usual equation for a massless gauge field, and the equation for the heavy HP is the same as the Proca equation we found for a massive photon, eq.~(\ref{mastereq}). We solve, therefore, in the same way as BD. The magnetic field associated to ${\bf{\tilde X}}$, given by ${\bf{B_{\tilde X}}}=\nabla \times {\bf{\tilde X}}$, is
\bb
{\bf{B_{\tilde X}}}=\hat z~e \tan\chi j \Theta(a-\rho)+\hat z m_{\gamma'}^2 \tan\chi  \Pi(\rho),
\ee
where the function $\Pi(\rho)$ is given again by eq.~(\ref{pieq}).

Now we can {\it go back} and find the true magnetic field,  ${\bf{B}}=\nabla \times {\bf{A}}$. From eq~(\ref{a1}) we have, ${\bf{A}}={\bf{\tilde A}}+\tan\chi{\bf{\tilde X}}$ and taking the curl to this equation we get
\bb
\nabla \times {\bf{A}}= \hat z~e j  \Theta(a-\rho)+\hat z~e \tan^2\chi j \Theta(a-\rho)+\hat z~m_{\gamma'}^2 \tan^2\chi  \Pi(\rho). \label{magneticb}
\ee
Therefore, we have realize that if a HP mixes with a photon, there is a small component of the magnetic field in a (confined) solenoid that leaks out of it, both inside and outside the solenoid radius. Let us note that therefore the effects stops being topological in nature, since there is actual leaking of magnetic field. This interesting fact opens several detection possibilities, besides Aharonov-Bohm-type experiments \cite{ABprep}.

\section{Limit from Aharonov-Bohm-type experiments}
In this section we will compute the bound that one can get from an AB experiment. Let us recall that the phase shift that an electron beam suffers by surrounding the solenoid is proportional to the magnetic flux enclosed by it
\bb
\Delta \varphi = e\Phi.
\ee
Without kinetic mixing, the magnetic flux is given by $\Phi_0= e j\pi a^2$. Assuming a mixing between photon and HP, $\Phi$ can be obtained by taking the surface integral of eq.(\ref{magneticb}), and reads
\bb
\Phi= \Phi_0\left(1+\tan^2\chi\right) + m_{\gamma'}^2 \tan^2\chi \int_S   \Pi(\rho) dS.
\label{flux}
\ee
The above formula is however not final yet, since it is not properly normalized. Note that for $m_{\gamma'} \rightarrow 0$, eq.~(\ref{flux}) does not recover its natural value, $\Phi_0$. This is because the electric charge gets also renormalized by the mixing photon-HP \cite{Endo:2012hp}.

This issue has been already addressed by Jaeckel and Roy in \cite{Jaeckel:2010xx}. They developed a procedure to get the proper bound from a function, $\mathcal F(\chi, m_{\gamma'},\alpha)$, where there is also a parameter $\alpha$ that depends on $\chi$. The procedure takes  as many independent measurements as the number of parameters  depend on the kinetic mixing in the function $\mathcal F$. In our case, we have $\chi$ and the electric charge, which we will write in terms of the fine structure constant $e^2= 4\pi \alpha$, so we need an extra independent measurement of $\alpha$ (besides the AB experiment). We will consider the electron $g-2$ experiment, worked on  \cite{Endo:2012hp}, since is the most sensitive measurement of $\alpha$. From ref.~\cite{Jaeckel:2010xx} we can write the bound on $\chi$ as
\bb
\chi^2 \leq \frac{\frac{|\Delta M_1|}{M_1}+\frac{|\Delta M_2|}{M_2}}{\left|\left( n_1 f_1(m_{\gamma'}) -n_2f_2(m_{\gamma'})\right)\right|},
\label{chibound}
\ee
where $M_i$, $(i=1,2)$, are the two independent measurements of $\alpha_i$, and have the form $M_i =c_i \alpha^{n_i} +\chi^2 f_i\left(m_{\gamma'}\right)$. The parameters $\Delta M_i$ are the absolute uncertainty in the measurement. The AB experiment provides a measurement on $\alpha$ given by
\bb
\Delta \varphi = 4\pi \alpha\tilde \Phi_0+4\pi \alpha  \chi^2 \tilde \Phi_0 f_1(m_{\gamma'}).
\ee
Where $\tilde \Phi_0=j\pi a^2$ and $f_1(m_{\gamma'})=1+\left(m_{\gamma'}^2/\Phi_0\right) \int_S \Pi(\rho)dS$.

\begin{figure}[t]
\centerline{\includegraphics[width=0.4\textwidth]{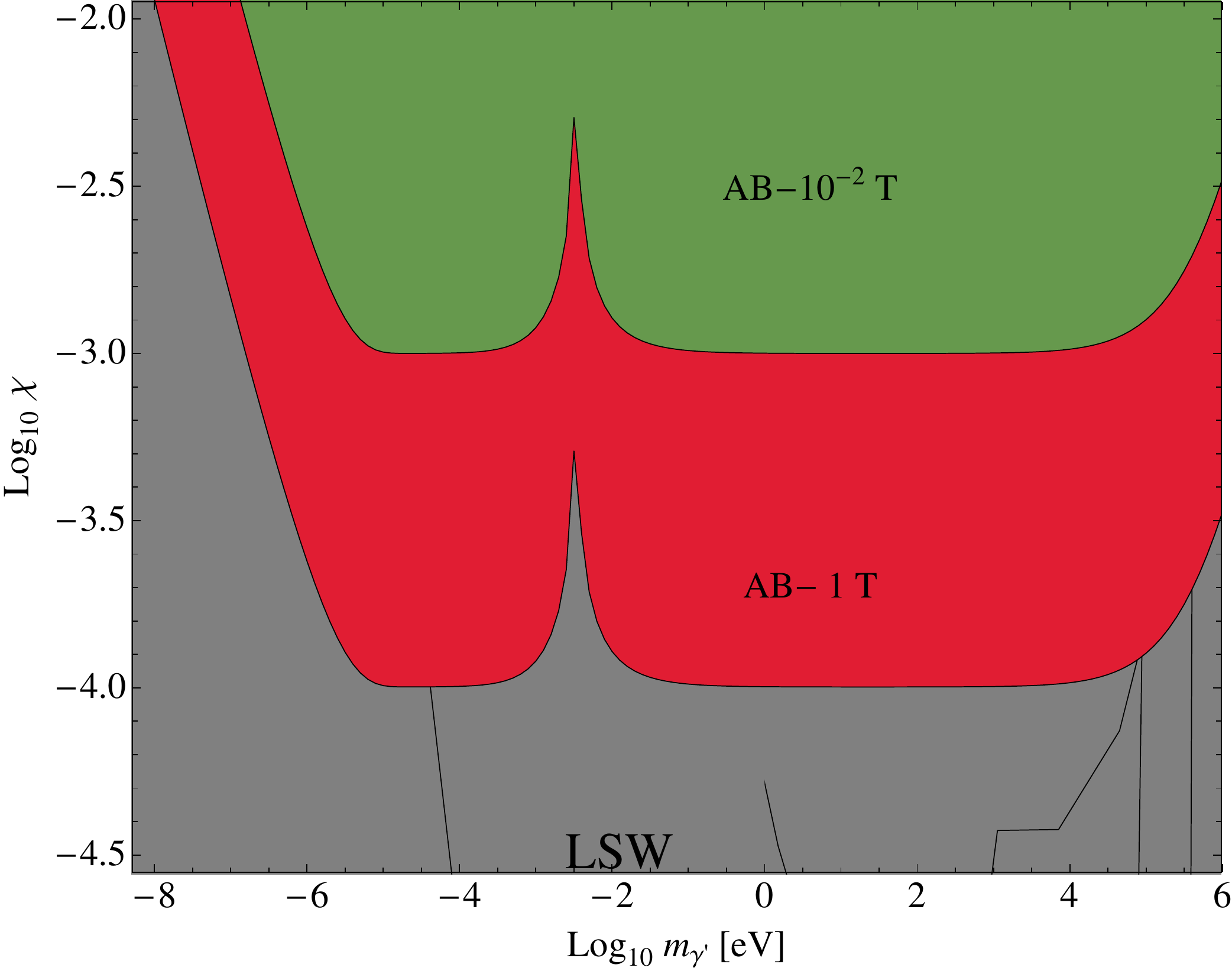}}
\caption{\footnotesize{Plot of kinetic mixing parameter ($\chi$) as a function of the HP mass ($m_{\gamma'}$).  The gray area corresponds to parameter space that has been already ruled out. For further explanation see text.}}
\label{fig:figure1}
\end{figure}

Figure~(\ref{fig:figure1}) shows our findings. We have considered two scenarios: the green bound considers a magnetic field of  $B=10^{-2}$ T and the red one is the optimistic scenario, considering a magnetic field of $B=$ 1~T. Both scenarios assume a sensitivity of the external magnetic field in the solenoid of $\Delta B= 10^{-8}~$T. The dimensions of the set up are a solenoid radius of $a=0.1~$cm  and an electron wavelength range of $\rho=10~$cm.

\section{Outlook}
We have explored the phenomenological consequences of mixing photons and hidden photons in the Aharonov-Bohm effect. We found that indeed there is a modification, and therefore the phase acquired by a test electron beam gets an additional shift in the phase.
The most attractive feature is to realize there is actual leaking of magnetic field out of the magnetic source. This implies that this AB effect -- known to be of topological nature -- gets an extra contribution to the phase shift of the electron beam that it has a non-topological nature.  
We also showed expected constrains from an hypothetical AB experiment. Further development will appear soon in \cite{ABprep}.
\section*{Acknowledgements}
I am very grateful of Andreas Ringwald, for valuable comments and encouragement, and the DESY Theory Group, for their hospitality. This work has been supported  by FONDECYT grant 11121403.
\begin{footnotesize}

\end{footnotesize}


\end{document}